\documentclass[sigconf,nonacm]{acmart}
\usepackage{amsmath}
\usepackage{array}
\usepackage{booktabs}

\acmYear{2026}
\copyrightyear{2026}

\begin{document}
\title{Concurrent Use of FP32 and FP64 Computing Units on GPUs and\\
Mixed-Precision Placement in WENO Reconstruction}

\author{Jun Hatayama}
\affiliation{\institution{Graduate School of Environmental, Life, Natural Science and
    Technology, Okayama University}
  \city{Okayama}
  \country{Japan}
}
\email{jhatayama@s.okayama-u.ac.jp}

\author{Daichi Mukunoki}
\affiliation{\institution{Information Technology Center, Nagoya University}
  \city{Nagoya}
  \country{Japan}
}
\email{mukunoki@cc.nagoya-u.ac.jp}

\author{Yusuke Onishi}
\affiliation{\institution{Graduate School of Environmental, Life, Natural Science and
    Technology, Okayama University}
  \city{Okayama}
  \country{Japan}
}
\email{p62d9qiy@s.okayama-u.ac.jp}

\author{Atsushi Sekimoto}
\affiliation{\institution{Faculty of Environmental, Life, Natural Science and
    Technology, Okayama University}
  \city{Okayama}
  \country{Japan}
}
\email{asekimoto@okayama-u.ac.jp}

\author{Toshinori Kouchi}
\affiliation{\institution{Faculty of Environmental, Life, Natural Science and
    Technology, Okayama University}
  \city{Okayama}
  \country{Japan}
}
\email{kouchi@okayama-u.ac.jp}
 
\begin{abstract}
Scientific computing, including computational fluid dynamics (CFD), often requires FP64 arithmetic, although FP64 performance is severely limited on consumer- and workstation-class GPUs, and the shift of GPU resources toward low-precision units for AI workloads may extend this limitation to data-center GPUs.
Double-float (DF) emulation, which represents a value using two FP32 values with about 48 bits of significand, outperforms native FP64 where FP64 units are limited. On the other hand, mixed-precision strategies offload to FP32 the operations that do not determine accuracy.
In both approaches, the native FP64 units remain idle during FP32 instruction execution because they are physically separate from the FP32 units.
We investigated whether this idle resource can be recovered on A100, GH200, GB200, and RTX PRO 6000 Blackwell Max-Q.
A microbenchmark shows that co-resident FP32 and native-FP64 kernels exceed the throughput of the faster kernel alone on A100 but not on GH200 and GB200; on RTX 6000 Blackwell, they do so against FFMA but, against DF emulation, only when the issue density of the DF kernel is reduced.
We then tested whether this gain carries over to computing the WENO-Z reconstruction, a standard high-order shock-capturing scheme in CFD, with the nonlinear weights in FP32 and the candidate polynomials in native FP64 (W32P64).
The weights are candidates for lower precision, although the precision they need depends on the order, problem, and implementation.
We observe substantial overlap on A100, whereas GH200, GB200, and RTX 6000 Blackwell show little overlap.
Importantly, the fastest placement does not always preserve the design accuracy of WENO-Z: for a one-dimensional entropy wave problem, W32P64 fails to achieve the design order of accuracy for WENO7 and WENO9.
In contrast, DF emulation of the entire reconstruction comes close to it on the grids tested for WENO5, WENO7, and WENO9 while delivering a 2.51--3.94$\times$ speedup over FP64 on RTX 6000 Blackwell.
For WENO5 on the two-dimensional Euler vortex problem, at the time step used, the W32P64 error at $n=256$ is 1.26 times that of FP64 and 1/104 of that of the all-FP32 reconstruction.
 \end{abstract}


\maketitle
\sloppy

\section{Introduction}
\label{sec:intro}

Driven by the growing demand from deep learning and AI workloads, 
GPU vendors have devoted an increasing share of compute resources to low-precision matrix units,
while FP64 capability has been steadily deprioritized, 
generation after generation.
On consumer- and workstation-class NVIDIA GPUs, FP64 throughput is only 1/64 of FP32 throughput. 
However, scientific computing, including computational fluid dynamics (CFD), often requires
FP64 arithmetic to capture high-fidelity physics.

Two approaches address this mismatch. 
One is mixed-precision computing, which reduces precision only in the operations that do not govern the accuracy of the
solution~\cite{kashi2026mixed,siklosi2025reduced,karp2025effects,chen2026enabling}.
The other is high-precision emulation, which represents a high-precision value with multiple low-precision numbers.
This paper focuses on double-word arithmetic~\cite{dekker1971floating,hida2001algorithms}, 
which represents a value as the unevaluated sum of two FP32 numbers, giving about 48 bits of significand. 
On GPUs with limited FP64 units, DF emulation can deliver several times the effective throughput of native FP64.

Under either approach, the same thing happens: while an FP32 instruction stream runs, the physically separate native FP64 units sit idle. 
This paper asks whether that idle resource can be recovered.
The ultimate form of recovery is intra-kernel: placing an FP32 instruction and an FP64 instruction side by side in one kernel and letting them overlap in time.
In a real compute kernel, however, whether overlap occurs depends on complex interactions among memory access, register pressure, and instruction dependencies. 
Consequently, when overlap fails to appear, it is difficult to determine whether hardware limitations or software bottlenecks are responsible.
We therefore begin with an inter-kernel microbenchmark that runs an FP32-only kernel and a native-FP64-only kernel in separate CUDA streams on the same SM and checks
whether the FP32 and FP64 pipes make progress concurrently. 
Warps from different kernels neither depend on each other nor share registers, so this experiment isolates the hardware-side condition: whether FP32 and FP64 instructions can be issued concurrently on one SM.
Our working hypothesis is that, on a GPU where inter-kernel overlap gives a throughput gain,
independent FP32 and FP64 instructions within one kernel may overlap as well.

As a CFD application, we choose the WENO-Z reconstruction~\cite{borges2008improved}, 
an improved variant of the weighted essentially non-oscillatory (WENO) scheme~\cite{liu1994weighted,jiang1996efficient},
which is a standard high-order shock-capturing scheme in CFD  
and is widely used for supersonic and hypersonic flows. 
The reconstruction builds candidate polynomials from local stencils and blends them with nonlinear weights based on smoothness indicators. 
The weights act as ratios that select among stencils and are therefore candidates for lower precision than the polynomial values, although the precision they need depends on the order, problem, and implementation.
The reconstruction thus splits naturally into two computations that can be given different precisions.
Field et al.~\cite{field2023gpu} exploited this split for black-hole simulations, computing the weights in FP64 and the polynomials in FP128. 
We apply the same split to recover idle hardware instead, computing the weights in FP32 and the polynomials in native FP64.
The WENO-Z kernel adds what the microbenchmark removed: its two computations share registers, read the same stencil values, and are ordered by the compiler, so any overlap must survive register pressure, occupancy and instruction order.

This paper treats mixed precision not only as a reduction in arithmetic precision but as a redistribution of work across distinct GPU execution resources, whose benefit depends jointly on the microarchitecture and on whether the scheme keeps its design order of accuracy.
 \section{Background and Design Model}
\label{sec:background}

\subsection{WENO-Z reconstruction and its decomposition}
\label{sec:weno}

We use the WENO-Z scheme~\cite{borges2008improved,castro2011high} and call its fifth-,
seventh-, and ninth-order variants WENO5, WENO7, and WENO9, respectively.
The reconstructed positive/negative flux components at the interface $i+1/2$ 
are the weighted sum of candidate polynomials $p_k$
and nonlinear weights $\omega_k$,
\begin{equation}
q_{i+1/2}^{\pm}=\sum_{k=0}^{r-1}\omega_k^{\pm}p_k^{\pm},
\label{eq:weno}
\end{equation}
where $r=3,4,5$ corresponds to WENO5, WENO7, and WENO9, and 
the superscripts $-$ and $+$ 
denote the left- and right-biased reconstructions, respectively. 
The candidate polynomials, $p_k^{\pm}$, and the smoothness indicators, $\beta_k$,
take the standard form~\cite{jiang1996efficient}: 
$\beta_k$ is a sum of squared derivatives of the $k$-th candidate polynomial 
and is a quadratic form in differences between stencil values. 
WENO-Z additionally uses a global smoothness indicator $\tau$,
\begin{equation}
\alpha_k=d_k\left\{1+\left(\frac{\tau}{\beta_k+\epsilon}\right)^{2}\right\},
\qquad
\omega_k=\frac{\alpha_k}{\sum_m\alpha_m},
\label{eq:weight}
\end{equation}
where $d_k$ are the linear weights, $\epsilon$ is a small positive constant, 
and $\tau=|\beta_0-\beta_2|$ for WENO5~\cite{borges2008improved} and the absolute value of a linear combination of the
$\beta_k$ for WENO7 and WENO9~\cite{castro2011high}. 
Hereafter we call the evaluation of $\beta_k$,
$\tau$, $\alpha_k$, and $\omega_k$, including the two divisions in Eq.~\eqref{eq:weight},
$\tau/(\beta_k+\epsilon)$ and $1/\sum_m\alpha_m$, the \emph{weight computation}; 
the evaluation of the candidate polynomials $p_k$, the \emph{polynomial computation}; 
and the final $\sum_k \omega_k p_k$, the \emph{blending}. 
The baseline mixed-precision placement, W32P64, 
computes the weights in FP32 and keeps the polynomial computation and the blending in FP64.

\subsection{Double-float emulation} \label{sec:df}

On GPUs with very low native FP64 throughput, W32P64 leaves the polynomial computation and the blending on the bottleneck FP64 pipe.
We therefore use double-float (DF) emulation, which achieves FP64-equivalent precision using
FP32 operations.
DF emulation represents a value as the unevaluated sum of two FP32 numbers, 
$a=a_{\rm hi}+a_{\rm lo}$. From the Dekker--Knuth error-free transformations~\cite{dekker1971floating,hida2001algorithms},
a DF addition takes 11 FP32 instructions, and a DF multiplication takes 7;
thus one DF multiply--add (two FP64-equivalent flops) takes 18 FP32 instructions.
Because the compiler's FMA contraction breaks the error-free transformations, 
we write addition, subtraction, and multiplication with rounding-mode-specified intrinsics, which the compiler does not contract. The resulting significand of approximately 48~bits is shorter than 
the 53 bits of FP64, and the exponent range remains that of FP32. 
Hereafter, ``FP64-equivalent'' refers to this precision and operation count; 
it does not guarantee the exponent range, subnormal handling, or exception behavior of IEEE~754 binary64.

\subsection{Instruction issue and FP32/FP64 execution lanes}
\label{sec:model}

An SM consists of four sub-partitions, and the warp scheduler of each sub-partition issues one
warp instruction per cycle~\cite{nvidia2020ampere,nvidia2022hopper,nvidia2025cudapg}. 
The number of FP32 and FP64 lanes per sub-partition is 16 and 8 for A100, 32 and 16 for GH200 (Hopper) and GB200 (data-center Blackwell), and 32 and 0.5 for RTX PRO 6000 Blackwell Max-Q (GB202; hereafter RTX 6000 Blackwell), respectively. 
With a warp width of 32, an FP32 warp instruction occupies the 16-lane pipe of A100 for two cycles and an FP64 warp instruction occupies the 8-lane pipe for four cycles.
An FP32 instruction stream on A100 therefore leaves every other issue slot free, 
and an FP64 stream needs even fewer slots. 
In terms of issue bandwidth, the two can share the scheduler. 
On GH200 and GB200, an FP32 stream can
issue every cycle, leaving, in principle, no slot for FP64 instructions. 
On RTX 6000 Blackwell, FP32 can also issue every cycle, 
but the FP64 pipe accepts only one instruction every 64 cycles,
so the two can share if the FP32 stream leaves as few as one issue slot in 64 unused.
This is why we expect A100 to behave differently from GH200 and GB200, despite their identical nominal FP64:FP32 ratio of 1:2.
We treat this expectation as a hypothesis: whether an instruction of the other pipe can
actually use a free slot depends on dispatch ports, operand collectors, register banks, and the
fairness of warp selection, which are not fully documented in publicly available NVIDIA materials. 
Lane counts therefore provide only a necessary condition, and Section~\ref{sec:slots} reports the
corresponding warp-scheduler counters.

\subsection{Inter-kernel and intra-kernel overlap}
\label{sec:levels}

FP32 and FP64 instructions can overlap in the following ways.
\emph{Inter-kernel overlap:} two kernels in separate CUDA streams co-reside on one SM, one issuing FP32 and the other FP64 instructions.
Because their warps neither depend on each other nor share registers, 
the requirement reduces to the hardware-side condition: both kernels hold SM resources simultaneously, and the warp scheduler must keep selecting warps from both kernels.
A stream that issues on nearly every cycle can leave no slots for the other, so the issue density of each stream, which we set 
through the length of its dependency chains, is part of this condition even though it is not a hardware property.

\emph{Intra-kernel overlap:} FP32 and FP64 instructions are issued within one kernel, and the overlap comes either from independent instructions within a thread (instruction-level parallelism, ILP) or from other warps (thread-level parallelism, TLP).
To yield ILP, the instructions of different types must be interleaved in the machine code of a thread (Section~\ref{sec:method-sass}).
On the other hand, TLP requires that the different types of instructions be issued from different warps.
Because both paths require free issue slots, we do not expect intra-kernel overlap on a GPU where inter-kernel overlap gives no throughput gain, and software-side conditions can prevent it.
 \section{Evaluation method}
\label{sec:method}

\subsection{Evaluation environment}
\label{sec:env}

Table~\ref{tab:env} lists the four GPUs. A100, GH200, and GB200 share the nominal FP64:FP32 ratio of 1:2. 
RTX 6000 Blackwell, with a measured ratio of 1:67.3 (nominal 1:64), is bound by native FP64, and therefore serves to evaluate DF emulation.
On GH200 and GB200 we also evaluated asynchronous stencil-tile transfer with the Tensor Memory Accelerator (TMA) (Section~\ref{sec:gh200}).

\begin{table*}[t]
\centering
\caption{GPUs used and their main characteristics. Measured peaks are the standalone values
(FP32~/~FP64) from the pure kernels of Section~\ref{sec:interpure}.}
\label{tab:env}
\begin{tabular}{lllll}
\toprule
 & A100 80GB PCIe & GH200 120GB & GB200 & RTX PRO 6000 Blackwell Max-Q \\
\midrule
Architecture / compute capability & Ampere / 8.0 & Hopper / 9.0 & Blackwell / 10.0 & Blackwell GB202 / 12.0 \\
SMs & 108 & 132 & 152 & 188 \\
FP32:FP64 lanes per SM & 64:32 & 128:64 & 128:64 & 128:2 \\
Measured peak [TFLOPS] (FP32 / FP64) & 19.10 / 9.424 & 52.63 / 31.679 & 65.40 / 38.696 & 99.80 / 1.483 \\
Measured FP64:FP32 & 1:2.0 & 1:1.66 & 1:1.69 & 1:67.3 \\
CUDA / driver & 12.5 / 570.211.01 & 13.0 / 595.58.03 & 13.3 / 610.57.04 & 13.2 / 595.58.03 \\
\bottomrule
\end{tabular}
 \end{table*}

\subsection{Inter-kernel microbenchmarks}
\label{sec:method-inter}

Three microbenchmarks share a common harness.
Each kernel launches one block per SM.
The two kernels are launched simultaneously from separate CUDA streams and timed with CUDA events, with the iteration count calibrated so that each kernel runs about 300~ms standalone. 
Each block records its SM ID and its start and end times using \texttt{clock64()}.
We define the retention of each kernel as its throughput during the interval in which the execution windows of the two kernels overlap on the same SM, divided by its standalone throughput.
The source code and compiler options are identical on the four GPUs except for the \texttt{-arch} option; the CUDA versions differ and are listed in Table~\ref{tab:env}.

\emph{Pure kernels (Section~\ref{sec:interpure}):} an FFMA kernel with eight independent accumulators per thread runs against a DFMA kernel of the same structure (25--28 registers per thread, no shared memory).
We sweep the FP32-side configuration and the FP64-side warp count.

\emph{DF emulation against native FP64 (Section~\ref{sec:interdf}):} the FP32 side is DF emulation, counted in FP64-equivalent flops (one DF operation as one flop), against a native FP64 kernel fixed at 1024 threads/SM (32 warps/SM), the configuration of its standalone peak.
Two configurations set the issue density of the DF side, both at 256 threads/SM: eight accumulators per thread,
which keeps eight independent chains, and two accumulators per thread, for which data dependencies leave
gaps in the issue stream.

\emph{Instruction-type probe (Section~\ref{sec:probe}):} this measurement repeats DFMA against a single instruction type such as FFMA, FADD, FMUL, or DF emulation, to determine which instruction types prevent overlap. Note that a DF multiply--add consists of 3 FFMA, 1 FMUL, and 14 FADD/FSUB instructions.

\subsection{WENO-Z kernel and static SASS analysis}
\label{sec:method-sass}

For CFD applications, we implement the WENO-Z reconstruction of Eqs.~\eqref{eq:weno}--\eqref{eq:weight} as a kernel.
In FP32, the two divisions in the weight computation use the approximate reciprocal, refined by one Newton step for the weight normalization, $1/\sum_m\alpha_m$.
FP64 keeps IEEE division: the FP32 approximate reciprocal (\texttt{MUFU.RCP}) is one instruction close to FP32 precision, whereas the FP64 one (\texttt{MUFU.RCP64H}) is only a seed that needs DFMA Newton steps in any case. \texttt{-use\_fast\_math} makes the same split.
Every placement on every GPU is measured with one hand-written CUDA C kernel, which the solvers of Section~\ref{sec:solver} also call, and each speedup is against its native FP64 in the same run.

We examine the streaming assembler (SASS), the native machine code in CUDA binaries, to determine the instruction order and mix without executing it (static analysis).
Because a warp issues the instructions of a thread in program order, a thread can supply overlap only where FP32 and FP64 instructions alternate; 
if they were grouped, overlap could come only from warps in different phases.
In the \texttt{sm\_80} (A100) SASS of the natural version $s_{\rm nat}$ of the kernel measured as $s_{\rm pure}$ (Section~\ref{sec:method-overlap}), 
the instruction class switches between FP32 and FP64 34 and 93 times at WENO5 and WENO9, respectively, compared with 3 times in its forced-serialization version.
In the \texttt{sm\_120} (RTX 6000 Blackwell) SASS, it switches 12 and 30 times.
The FP64 instruction counts of the \texttt{sm\_80}, \texttt{sm\_90} (GH200), \texttt{sm\_100} (GB200) and \texttt{sm\_120} binaries agree exactly (61, 127, and 187 for WENO5, WENO7, and WENO9), 
the FP32 counts agree to within 2.3\%, and no register spills occur in the measured kernels; the out-of-line routine of the FP64 IEEE division spills 4~bytes on \texttt{sm\_90} and \texttt{sm\_120}.
Timing differences between generations are therefore not explained by a large difference in instruction counts.

\subsection{Measuring overlap in the WENO-Z kernel}
\label{sec:method-overlap}

In an ordinary kernel, memory access and arithmetic proceed concurrently, and 
the run time alone cannot separate their contributions. 
We therefore use a compute-repeat configuration: the stencil input is read from global memory once, 
the reconstruction is repeated $N$ times on the same input, 
and the result is written back once. 
The slope of a least-squares linear fit of run time against $N$ gives the arithmetic time per reconstruction, excluding memory access. 
Let $s_{32}$ and $s_{64}$ denote the slope of the weight computation alone and the polynomial computation alone, respectively, and $s_{\rm pure}$ 
the slope of the kernel that performs both computations.
We define the overlap indicator as
\begin{equation}
E=\frac{s_{32}+s_{64}-s_{\rm pure}}{\min(s_{32},s_{64})}.
\label{eq:overlap}
\end{equation}
The numerator is the time saving relative to serial execution, $s_{32}+s_{64}$, 
and the denominator is its value when the shorter stream is fully hidden; $E=0$ corresponds to fully serial execution, and $E<0$ to a kernel slower than serial execution.
Because the partial kernels may differ from the combined kernel in register usage and compiler optimization, $E$ is an indicator of overlap, not a measurement of the overlap time.
When one stream is much longer than the other, the denominator is small, and a timing change of a few percent of the longer stream moves $E$ by a large amount;
we therefore report $s_{64}/s_{32}$ alongside $E$.

A positive $E$ does not distinguish intra-thread ILP from TLP supplied by other warps. 
To separate them, we compare against a forced-serialization kernel that removes only the intra-thread overlap while matching the instruction count, bitwise results, and occupancy. 
In the serialized version (slope $s_{\rm ser}$), the input of the polynomial computation depends on the output of the weight computation; 
the independent version (slope $s_{\rm ind}$) adds the same dependent instructions but 
applies them to an input that is already final.
The added instructions appear in both versions, so their cost cancels in the difference $s_{\rm ser}-s_{\rm ind}$, the only quantity we use. \section{Microbenchmark: concurrent FP32/FP64 use across kernels}
\label{sec:inter}

\subsection{Pure FFMA/DFMA kernels}
\label{sec:interpure}

We first measure pure compute kernels, which most directly test the expectation of Section~\ref{sec:model}.
Across the sweep of the FP32-side configuration and the FP64-side warp count, 
blocks of both kernels co-resided on every SM of all four GPUs, and their execution windows overlapped completely.

Over 128--1024 FP32 threads and 4--16 FP64 warps per SM, on A100, the FP32 side retains 89--92\% and the FP64 side 77--89\%, saturating at 12~warps/SM.
In the best configuration (256 FP32 threads, 384 FP64 threads per SM), the two kernels sustain 90.6\% and 89.1\%, respectively,
and the wall time is 44.0\% shorter than when the two kernels run back-to-back.
On RTX 6000 Blackwell, FP64 retention likewise recovers as the FP64 warp count increases, from 40.2\% at 4 warps to 94.8\% at 16, and the best configuration shortens the wall time by 47.1\%.
GH200 behaves differently: over the same range, FP64 retention remains at only 15--65\%, while the FP32 side keeps 79--97\%, and even the best configuration (128 FP32 threads, 128 FP64 threads per SM) reduces the wall time by
only 27.9\%. 
GB200 behaves like GH200: over the same range, FP64 retention stays at 18--37\% while the FP32 side keeps 89--92\%, and the best configuration, the same as GH200's, reduces the wall time by 15.8\%.
This pattern agrees with the lane-count expectation of Section~\ref{sec:model}.

\begin{table*}[t] \centering
\caption{Concurrent execution of DF emulation and native FP64 in two streams, in FP64-equivalent TFLOPS.
The standalone columns give each kernel run alone in the same configuration.
The DF side and FP64 side
columns are the throughput during the overlap interval, with the retention
relative to standalone in parentheses. The last column is the combined throughput divided by the
largest standalone value of the GPU; a configuration below 1 is faster without concurrency.}
\label{tab:stream-df}
\begin{tabular}{llrrrrrr}
\toprule
 & & \multicolumn{2}{c}{Standalone [TFLOPS]} & \multicolumn{3}{c}{Concurrent [TFLOPS]} & \\
\cmidrule(lr){3-4}\cmidrule(lr){5-7}
GPU & Configuration & DF & FP64 & DF side & FP64 side & Combined & vs.\ faster standalone \\
\midrule
A100 & 8 accumulators & 1.071 & 9.441 & 0.943 (88.0\%) & 8.852 (93.8\%) & 9.795 & 1.038 \\
     & 2 accumulators & 1.071 & 9.441 & 0.933 (87.1\%) & 8.686 (92.0\%) & 9.619 & 1.019 \\
\midrule
GH200 & 8 accumulators & 3.03 & 31.43 & 1.74 (57.3\%) & 17.07 (54.3\%) & 18.80 & 0.598 \\
      & 2 accumulators & 3.38 & 31.43 & 2.14 (63.3\%) & 14.66 (46.6\%) & 16.80 & 0.535 \\
\midrule
GB200 & 8 accumulators & 4.130 & 38.695 & 3.344 (81.0\%) & 8.392 (21.7\%) & 11.736 & 0.303 \\
      & 2 accumulators & 4.105 & 38.695 & 2.645 (64.4\%) & 17.619 (45.5\%) & 20.264 & 0.524 \\
\midrule
RTX 6000 & 8 accumulators & 5.06 & 1.49 & 5.02 (99.2\%) & $\sim$0 (starved) & 5.03 & 0.994 \\
          & 2 accumulators & 4.80 & 1.49 & 4.18 (87.2\%) & 1.215 (81.6\%) & 5.40 & 1.067 \\
\bottomrule
\end{tabular}
 \end{table*}

\subsection{DF emulation and native FP64}
\label{sec:interdf}

Table~\ref{tab:stream-df} gives the standalone throughput of each kernel and the results of concurrent execution in two streams.
On RTX 6000 Blackwell, the faster DF configuration achieves 3.4 times the throughput of native FP64, whereas on A100, GH200, and GB200,
native FP64 achieves about nine times the throughput of DF emulation. 

On RTX 6000 Blackwell, when native FP64 runs concurrently with the ordinary 
DF kernel with eight accumulators per thread, the DF emulation side retains 99.2\%, while
the FP64 side almost stops during the overlap. 
We call this state \emph{starvation}: DFMA instructions obtain almost no issue opportunities, and their effective throughput falls to nearly zero.
Reducing the accumulator count from eight to two, so that data dependencies leave natural gaps in the issue stream of the DF kernel, 
allows DF emulation and FP64 to retain 87.2\% and 81.6\%, respectively, reaching 5.40 FP64-equivalent TFLOPS,
$6.7$\% above DF emulation alone.

On A100, both configurations retain 87--88\% on the DF side and 92--94\% on the FP64 side, and the eight-accumulator configuration performs best
at 9.795 TFLOPS, 3.8\% above the 9.441 TFLOPS of native FP64 alone.
Concurrency itself thus works well. 
However, DF emulation alone reaches only about 1/9 of native
FP64 throughput, so adding independent DF work yields only a small gain.
GH200, despite an FP64:FP32 ratio close to that of A100, drops to 47--63\% retention on both sides, and in every configuration the combined throughput falls 
at least 40.2\% below that of native FP64 alone.
GB200 falls further, to 30.3\% and 52.4\% of native FP64 alone with eight and two accumulators per thread.

Without interference, the gain over the faster standalone kernel is bounded by
the throughput ratio of the slower kernel to the faster one: about 29\% on RTX 6000 Blackwell (6.7\% measured), 11.3\% on A100 (3.8\% measured), 10.8\% on GH200, and 10.7\% on GB200 (negative measured gain on both). 
Gains must therefore be judged against the faster standalone kernel, not against co-residency or wall-time reduction.
We refer to a combined throughput exceeding that of either standalone kernel as an inter-kernel throughput gain, distinct from temporal overlap.

\subsection{Instruction-type probe}
\label{sec:probe}

The essential difference between the DF emulation kernel and the pure FFMA kernel is the instruction mix. We therefore
isolated its effect by changing only the instruction type of the competing kernel (Table~\ref{tab:probe}).

\begin{table}[t] \centering
\caption{Retention of the native-FP64 kernel (32 warps/SM) during the overlap interval with a competing kernel (256 threads/SM). 
}
\label{tab:probe}
\setlength{\tabcolsep}{3.5pt}
\begin{tabular}{lrrrr}
\toprule
Competing stream & A100 & GH200 & GB200 & RTX 6000 \\
\midrule
Continuous FFMA & 89.0\% & 20.3\% & 18.8\% & 93.2\% \\
Continuous FADD & 90.7\% & 0.4\% & 0.3\% & 0.4\% \\
Continuous FMUL & 89.4\% & 0.1\% & 0.3\% & 0.4\% \\
Alternating FMUL/FADD & 88.3\% & 0.3\% & 0.3\% & 0.4\% \\
DF (ordinary) & 93.0\% & 54.1\% & 41.4\% & 0.6\% \\
DF (all-FMA) & 89.4\% & 51.0\% & 40.9\% & 75.4\%$^\dagger$ \\
Low-density FADD (1 acc.) & 87.2\% & 67.9\% & 71.3\% & 100.1\% \\
\bottomrule
\end{tabular}
 \end{table}

On RTX 6000 Blackwell, GH200, and GB200, saturated single-type FADD or FMUL streams block DFMA almost completely (FP64 retention of 0.1--0.4\%). 
On A100, the same probes keep FP64 retention at 87--93\% for every instruction type, and no starvation is observed; 
with the competing kernel fixed, varying the FP64 warp count from 8 to 32 changes the retention by only 1--2 points. 

On RTX 6000 Blackwell, FADD blocks DFMA but FFMA does not (0.4\% against 93.2\%).
Because 14 of the 18 instructions of a DF multiply--add are FADD/FSUB, this is consistent with the starvation of native FP64 by DF emulation, but not by FFMA, on that GPU.

We also rewrote every addition and multiplication of the DF kernel as an FMA,
which preserves rounding and instruction count. 
On RTX 6000 Blackwell, this rewrite raised FP64 retention to 75.4\% (marked $\dagger$ in Table~\ref{tab:probe}).
The same transformation embedded in a different program, however, starved the FP64 kernel completely again, and on A100 and GH200 it lowered the retention, while on GB200 it changed it by only 0.5 points.
The rewrite is therefore not a portable remedy.

In summary, an inter-kernel throughput gain appears on A100 but not on GH200 and GB200.
On RTX 6000 Blackwell, it appears against FFMA but, against DF emulation, only when the issue density of the DF kernel is reduced. By the logic of Section~\ref{sec:levels},
intra-kernel overlap in the WENO-Z kernel can therefore be expected on A100, not on GH200 or GB200, and on RTX 6000 Blackwell only conditionally. 
 \begin{table*}[t]
\centering
\caption{Run time per precision placement on the four GPUs, speedup over FP64, and the overlap
indicator. $E$ is measured with the compute-repeat configuration of
Section~\ref{sec:method-overlap}, at WENO5 and WENO9 only. Each parenthesis is that row's FP64 cell over the cell's $t_{\min}$. $\dagger$:
$s_{64}/s_{32}=13.8$ is extremely asymmetric and $E$ is not evidence of overlap. 
DF emulation on A100, GH200, and GB200 is out of scope.}
\label{tab:overview}
\begin{tabular}{lcrrrr}
\toprule
 & & \multicolumn{3}{c}{$t_{\min}$ [$\mu$s] (speedup over FP64 in parentheses)} & \\
\cmidrule(lr){3-5}
GPU & WENO order & FP64 & W32P64 & DF & $E$ \\
\midrule
A100 & 5 & 606.2 & 199.7 (3.04) & -- & $0.89$ \\
A100 & 7 & 739.3 & 270.3 (2.73) & -- & -- \\
A100 & 9 & 1078.3 & 392.2 (2.75) & -- & $0.84$ \\
\midrule
GH200 & 5 & 264.2 & 95.1 (2.78) & -- & $0.12$ \\
GH200 & 7 & 269.7 & 138.9 (1.94) & -- & -- \\
GH200 & 9 & 431.3 & 202.8 (2.13) & -- & $-0.02$ \\
\midrule
GB200 & 5 & 220.3 & 76.4 (2.88) & -- & $0.03$ \\
GB200 & 7 & 233.1 & 120.5 (1.93) & -- & -- \\
GB200 & 9 & 370.4 & 174.2 (2.13) & -- & $-0.13$ \\
\midrule
RTX 6000 & 5 & 3303.1 & 557.5 (5.92) & 838.0 (3.94) & $1.00^\dagger$ \\
RTX 6000 & 7 & 4563.6 & 1021.3 (4.47) & 1629.3 (2.80) & -- \\
RTX 6000 & 9 & 6773.9 & 1415.6 (4.79) & 2694.5 (2.51) & $-0.45$ \\
\bottomrule
\end{tabular}
 \end{table*}

\section{Mixed-precision reconstruction inside the WENO-Z kernel}
\label{sec:intra}

This section implements the reconstruction of Section~\ref{sec:weno} as a single kernel and measures three precision placements, FP64, W32P64, and DF emulation.
We measure DF emulation only on RTX 6000 Blackwell, because on A100, GH200, and GB200 native FP64 achieves about nine times the throughput of DF emulation (Section~\ref{sec:interdf}).

Table~\ref{tab:overview} highlights two key points. 
First, W32P64 is faster than FP64 on all GPUs, with speedups of 2.73--3.04$\times$, 1.94--2.78$\times$, and 1.93--2.88$\times$ on A100, GH200, and GB200, respectively.
With IEEE division in the FP32 weights, they drop to 1.82--2.25$\times$, 1.35--1.59$\times$, and 1.29--1.62$\times$.
On RTX 6000 Blackwell, W32P64 gives the largest speedup, 4.47--5.92$\times$ (4.54--5.89$\times$ with IEEE division),
but, as Section~\ref{sec:solver} shows, it loses the design order of accuracy for WENO7 and WENO9; 
the fastest placement that preserves the design order is DF emulation, 
with a speedup of 2.51--3.94$\times$. 
These speedups also include the changes in arithmetic and division cost and in register usage, so they do not isolate the contribution of overlap.
Second, the temporal overlap of the arithmetic depends strongly on the GPU generation, and this tendency is consistent with the results of the microbenchmarks in Section~\ref{sec:inter}.

\subsection{A100: overlap and its source}
\label{sec:a100}

For WENO9,
$s_{32}=115.84$, $s_{64}=60.59$, and $s_{\rm pure}=125.49~\mu$s/iteration, giving $E=0.84$, and $E=0.89$ for WENO5, at $s_{64}/s_{32}=0.52$.
These values and the Nsight Compute counters are consistent with most of the shorter FP64 polynomial computation overlapping the FP32 weight computation:
in the mixed kernel, FP64 pipe utilization falls from 98--99\% standalone to 48--49\%, 
while the FP32 FMA pipe utilization falls only from 98\% to 91--93\%.

As expected from Section~\ref{sec:model}, the FP32 stream on A100 leaves issue slots free (41--46\%, Table~\ref{tab:slots}), 
which is the precondition for the overlap. 
Whether subsequent instructions from the same thread or other warps fill these slots 
cannot be determined from this measurement alone; we therefore ran the forced-serialization control in Section~\ref{sec:method-overlap} (Table~\ref{tab:serial}).

\begin{table*}[t]
\centering
\caption{Forced-serialization control on A100, $s$ in $\mu$s/iteration, with $s_{\rm null}=s_{32}+s_{64}$ the zero-overlap reference.
$E_{\rm nat}$ comes from the natural version $s_{\rm nat}$ at fixed occupancy and differs from $E$ in Table~\ref{tab:overview} because $s_{\rm nat}$ is an occupancy-pinned copy of the kernel (0.85 against 0.84 at WENO9).
``dependent'' joins the two streams through the blending once per reconstruction, so $s_{\rm null}$ and $E_{\rm nat}$ are undefined for it.
$s_{\rm ser}$ and $s_{\rm ind}$ match in static and executed instruction count, occupancy, and busy time of both pipes.}
\label{tab:serial}
\begin{tabular}{llrrrrrrl}
\toprule
WENO order & Placement & $s_{\rm nat}$ & $s_{\rm ser}$ & $s_{\rm ind}$ & $s_{\rm null}$ & $E_{\rm nat}$ & $s_{\rm ser}/s_{\rm ind}-1$ [\%] & Outcome \\
\midrule
WENO5 & independent & 42.19 & 44.16 & 44.05 & 60.71 & 0.89 & +0.24 & serial slower \\
WENO5 & dependent & 46.41 & 47.83 & 47.71 & -- & -- & +0.24 & within noise \\
WENO9 & independent & 124.80 & 124.23 & 125.82 & 176.48 & 0.85 & $-1.27$ & serial faster \\
WENO9 & dependent & 129.81 & 131.95 & 132.58 & -- & -- & $-0.48$ & serial faster \\
\bottomrule
\end{tabular}
 \end{table*}

The natural version is 29--31\% faster than the zero-overlap reference, $s_{\rm null}=s_{32}+s_{64}$, which is consistent with temporal overlap.
Forced serialization, however, recovers none of that gap: the fraction returned toward
$s_{\rm null}$, $(s_{\rm ser}-s_{\rm ind})/(s_{\rm null}-s_{\rm ind})$, 
is only $+0.01$ for WENO5 and $-0.03$ for WENO9, and the serialization changes the run time by $+0.24$\% for WENO5 and $-1.27$\% for WENO9. 
This suggests that TLP sustains the overlap on A100, but does not rule out intra-thread ILP.
We took this measurement at 9.8 (WENO5) and 5.9 (WENO9) resident warps per scheduler.

\subsection{GH200 and GB200: loss of overlap}
\label{sec:gh200}

The same compute-repeat configuration gives the opposite result on GH200. 
$E$ is 0.12 for WENO5 and $-0.02$ for WENO9, at $s_{64}/s_{32}=0.44$: the FP32 weight computation and the FP64 polynomial computation execute almost serially. 
For WENO9, $s_{\rm pure}=59.87~\mu$s/iteration against
$s_{32}+s_{64}=59.58~\mu$s/iteration, which agree within 0.5\%, and Nsight Compute reports an FP64 pipe utilization of 29\%, against 29.1\% expected for fully serial execution. 
Increasing the FP64-side work did not restore simultaneous issue. 
These observations are consistent with the expectation of Section~\ref{sec:model}: an FP32 instruction stream leaves GH200 almost no free issue slots, and FP64 instructions therefore extend the critical path instead of filling gaps.
The FP64 retention remained low in Section~\ref{sec:interpure}, even as more FP64 warps were added, supporting the same interpretation.

GB200, which has the lane counts of GH200 (Section~\ref{sec:model}), gives the same result: $E=0.03$ for WENO5 and $-0.13$ for WENO9, at $s_{64}/s_{32}=0.43$--0.44.
For WENO9, $s_{\rm pure}=52.02~\mu$s/iteration is 3.9\% above $s_{32}+s_{64}=50.08~\mu$s/iteration, and the FP64 pipe utilization of the mixed kernel is 28.0\%, against 28.8\% for fully serial execution.
The loss of overlap on GH200 is therefore not specific to Hopper: it occurs on a later generation with the same lane counts, as does the low FP64 retention of Section~\ref{sec:interpure}.

TMA-based asynchronous stencil-tile transfer, in contrast, was effective on GH200: it was 4.0\%, 1.5\%, and 3.3\% faster than the non-tiled version for WENO5, WENO7, and WENO9, and 10.1\%, 5.0\%, and 4.3\% faster than filling shared memory manually in each thread; on GB200 the corresponding gains are 3.8\%, 4.7\%, and 5.2\%, and 10.2\%, 10.7\%, and 12.6\%. 

\subsection{RTX 6000 Blackwell: moving off the bottleneck pipe}
\label{sec:blackwell}

On RTX 6000 Blackwell, the native FP64 throughput is about 1/67.3 of FP32, and the FP64 WENO-Z kernel is bound by the FP64 pipe (FP64 pipe utilization 86.4--86.9\%, issue-slot utilization 7.1--8.2\%). 
Applying the DF emulation of Section~\ref{sec:df} to the whole reconstruction
gives speedups of 3.94$\times$, 2.80$\times$, and 2.51$\times$ over the FP64 baseline 
for WENO5, WENO7, and WENO9. 
A placement that converts only the weight computation to DF emulation and leaves the polynomial computation and the blending in native FP64 achieves a speedup of only 2.39--3.05$\times$.
Although most FP64 instructions are in the weight computation, 
converting it alone leaves the remaining FP64 instructions on the bottleneck pipe, and among the DF placements measured, 
the speedup is determined by moving these instructions to the FP32 pipe as well.

This GPU shows even more clearly that free issue capacity is not a sufficient condition for overlap. For W32P64, $E=-0.45$ for WENO9: 
no overlap appears, even though
the run time of the mixed kernel is 7.8--8.9$\times$ the issue-slot demand of the two streams combined, so issue bandwidth does not limit it.
At WENO9, adding the FP32 instructions raises the register count from 58 to 72 and lowers occupancy from 65.5\% to 57.5\%; register usage and the number of eligible warps may limit the overlap.
The inter-kernel throughput gain of Section~\ref{sec:interdf} at reduced issue density does not carry over to one kernel.

For DF emulation, we report only the 2.51--3.94$\times$ speedup and make no claim about temporal overlap within it.

\subsection{Warp-scheduler counters}
\label{sec:slots}

\begin{table}[t]
\centering
\setlength{\tabcolsep}{4.5pt}
\caption{Warp-scheduler counters for the weight computation alone (FP32), the polynomial computation alone (FP64), and both in one kernel (mixed), in the compute-repeat configuration.
Issue-slot utilization is the fraction of cycles on which a sub-partition's scheduler issued.
In the mixed kernel, ``eligible'' is warps eligible per scheduler per active cycle and ``not sel.'' warps eligible but not selected per issue-active cycle.}
\label{tab:slots}
{\small \begin{tabular}{llrrrrr}
\toprule
 & & \multicolumn{3}{c}{issue-slot utilization [\%]} & \multicolumn{2}{c}{mixed kernel} \\
\cmidrule(lr){3-5}\cmidrule(lr){6-7}
GPU & WENO & FP32 & FP64 & mixed & eligible & not sel. \\
\midrule
A100 & 5 & 58.8 & 28.7 & 67.7 & 4.56 & 5.71 \\
A100 & 9 & 54.4 & 26.3 & 62.1 & 2.75 & 3.41 \\
\midrule
GH200 & 5 & 88.8 & 72.4 & 84.9 & 6.98 & 7.18 \\
GH200 & 9 & 87.7 & 69.9 & 81.3 & 4.22 & 4.17 \\
\midrule
GB200 & 5 & 88.1 & 71.6 & 82.9 & 6.82 & 7.18 \\
GB200 & 9 & 86.2 & 68.8 & 79.3 & 4.68 & 4.87 \\
\midrule
RTX 6000 & 5 & 91.6 & 6.8 & 11.9 & 0.25 & 1.11 \\
RTX 6000 & 9 & 88.1 & 6.8 & 11.2 & 0.23 & 1.06 \\
\bottomrule
\end{tabular}
 }
\end{table}

Table~\ref{tab:slots} tests the lane-count expectation of Section~\ref{sec:model} directly.
The FP32 stream alone leaves 41--46\% of the issue slots free on A100, 11--12\% on GH200, and 12--14\% on GB200, the order the lane counts suggest.
Adding the FP64 stream raises the issue rate on A100 and lowers it on GH200 and GB200.
The GH200 and GB200 mixed kernels also have more eligible warps per scheduler than the A100 mixed kernel. Thus, a shortage of eligible warps cannot explain the difference between these GPUs.

These counters are consistent with the reading of Section~\ref{sec:model} but do not establish it.
RTX 6000 Blackwell shows their limit: its schedulers issue on only 11--12\% of cycles, so issue bandwidth does not limit that GPU, and its mixed kernel has almost no eligible warps (0.23--0.25 per scheduler). These counters do not show what decides whether one pipe can use a slot the other leaves, and we do not claim to have identified it.
 \section{Numerical Accuracy at the Solver Level}
\label{sec:solver}

The speedups of Section~\ref{sec:intra} do not decide whether a placement is usable.
Each placement was therefore embedded in an Euler solver to test whether it preserves the design order of the scheme, the order of accuracy it is constructed to attain on smooth solutions: 5, 7, and 9 for WENO5, WENO7, and WENO9.
Against FP64 we compare DF emulation (labeled DF in Table~\ref{tab:overview} and in the figures), W32P64, and all-FP32.
We tested two smooth fields, an entropy wave and an Euler vortex, because the computation of the smoothness indicator is sensitive to reduced precision, and our objective is to assess this sensitivity. Accordingly, we make no claims about shock-capturing performance.

Both fields are governed by the two-dimensional compressible Euler equations in conservative form,
\begin{equation}
\frac{\partial\mathbf{Q}}{\partial t}+\frac{\partial\mathbf{E}}{\partial x}
+\frac{\partial\mathbf{F}}{\partial y}=\mathbf{0},
\label{eq:euler}
\end{equation}
with $\mathbf{Q}=(\rho,\rho u,\rho v,e)^{\top}$, $\mathbf{E}=(\rho u,\rho u^{2}+p,\rho uv,(e+p)u)^{\top}$, and $\mathbf{F}=(\rho v,\rho uv,\rho v^{2}+p,(e+p)v)^{\top}$.
Here $\rho$ is the density, $(u,v)$ the velocity, $p$ the pressure, and $e=p/(\gamma-1)+\rho(u^{2}+v^{2})/2$ the total energy per unit volume of an ideal gas with $\gamma=1.4$; for the one-dimensional field, $v=0$ and nothing depends on $y$.
The one-dimensional field is computed by a one-dimensional solver and the vortex by a two-dimensional one.
The scheme is a nodal finite-difference WENO-Z scheme with global Lax--Friedrichs flux splitting, where the WENO reconstruction is performed independently in each coordinate direction using one-dimensional operators.

Both fields are non-dimensionalized with the domain length $L$, the background density $\rho_0$, and the speed of sound $c_0$ (pressure by $\rho_0c_0^{2}$, time by $L/c_0$) and are carried by a uniform free stream at the Mach number $M_0=0.05$.
Time integration uses the classical fourth-order Runge--Kutta method with $\Delta t=0.03\,h/M_0$, where $h=1/n$ is the spacing of a grid with $n$ points in each direction in which the field varies.
This is a convective CFL number $u\Delta t/h$ of 0.03 and an acoustic CFL number $(u+c)\Delta t/h$ of 0.63, with $c$ the speed of sound.
Both fields are periodic and are advected over exactly one period, $t_f=1/M_0$, after which the exact solution is the initial field.
Errors are reported as $L_2(\rho-\rho_{\rm exact})/\rho_0$, the root-mean-square nodal density error.

\subsection{Design order on a one-dimensional entropy wave}
\label{sec:entropy}

\begin{figure}[t]
\centering
\includegraphics[width=\columnwidth]{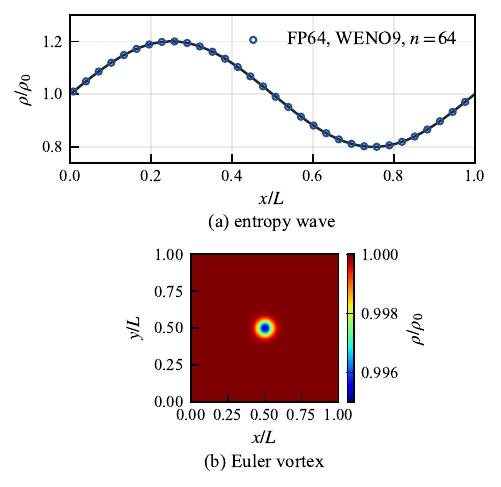}
\caption{The two smooth flow fields on which the placements are compared, as the density normalized by the background density.
(a)~The entropy wave: the line is Eq.~\eqref{eq:entropy-ic}, the initial field and, by periodicity, the exact solution after one period; the circles are the computed FP64 solution after one period at WENO9, $n=64$.
(b)~The Euler vortex of Eq.~\eqref{eq:evc-ic} after one period, computed in FP64 at WENO5, $n=256$, $\beta_{\rm v}=2$.}
\label{fig:cases}
\end{figure}

The first test case is a one-dimensional entropy wave.
On the periodic domain $x\in[0,1)$ the initial condition is
\begin{equation}
\frac{\rho(x,0)}{\rho_0}=1+A\sin(2\pi x),\quad A=0.2,
\qquad (u,p)=\left(M_0,\tfrac{1}{\gamma}\right).
\label{eq:entropy-ic}
\end{equation}
The velocity and the pressure are uniform, so the density perturbation is advected without change of shape (Figure~\ref{fig:cases}(a)).
The grids are $n=16,32,64$; on finer grids the truncation error falls below the noise floor of DF emulation.

\begin{figure*}[t]
\centering
\includegraphics[width=\textwidth]{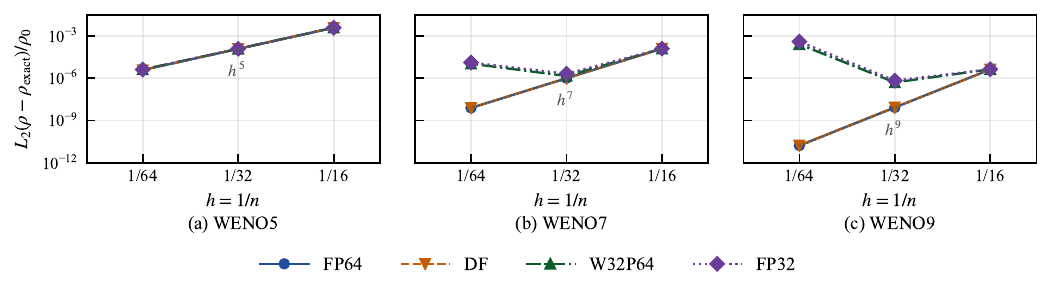}
\caption{Grid convergence on the one-dimensional entropy wave at (a)~WENO5, (b)~WENO7, and (c)~WENO9: the normalized density error $L_2(\rho-\rho_{\rm exact})/\rho_0$ after one period against the grid spacing $h=1/n$.
Dotted lines show the slope of the design order.}
\label{fig:entropy-conv}
\end{figure*}

Figure~\ref{fig:entropy-conv} shows the results, with the fitted order defined as the least-squares slope of $\log L_2$ against $\log h$ over $n=16,32,64$.
FP64 comes close to the design orders on these grids, with 4.98/6.98/8.96 at WENO5/7/9, and DF emulation follows at 4.98/6.98/8.95.
At $n=64$ the errors of the two are identical at WENO5, agree to five significant digits at WENO7, and differ by a factor of 1.01 at WENO9 ($1.690\times10^{-11}$ against $1.669\times10^{-11}$).
We attribute this difference to the noise floor of the roughly 48-bit DF significand, not to a loss of design order.
The FP32-weight placements, in contrast, keep the design order only at WENO5, with a fitted order of 4.91 for both and $n=64$ errors within a factor of 1.10 of FP64.
At WENO7 the fitted orders of W32P64 and all-FP32 fall to 1.78 and 1.64, and at WENO9 to $-2.94$ and $-3.26$.
At both orders the error increases from $n=32$ to 64; at WENO9, that of W32P64 grows from $4.97\times10^{-7}$ to $2.52\times10^{-4}$, seven orders of magnitude above the $1.67\times10^{-11}$ of FP64.

This loss of order may be attributed to the smoothness indicators.
The $\beta_k$ of WENO7/9 are quadratic forms with many cross terms, which can cancel more strongly as refinement makes the field locally smoother.
In FP32, $\beta_k$ may lose its significant digits while the truncation error is still falling, so $\tau/(\beta_k+\epsilon)$ in Eq.~\eqref{eq:weight} can be dominated by round-off.
Consequently, in our implementation and test conditions, W32P64, the placement with the largest speedup on RTX 6000 Blackwell in Section~\ref{sec:intra}, does not preserve the design order at WENO7 and WENO9.
Of the three reduced-precision placements tested, the only one that stays close to the design order there on these grids is DF emulation.

\subsection{Grid convergence on Euler vortex advection}
\label{sec:evc}

We next test WENO5 using a two-dimensional field, an advected isentropic vortex; WENO7/9 are not evaluated, because the FP32-weight placements already lose the design order there in a one-dimensional problem.
On the periodic domain $[0,1)^2$ a vortex of core radius $R_c=0.05$ is superimposed on the uniform flow $(u,v,p)=(M_0,0,1/\gamma)$,
\begin{equation}
\frac{u_\theta}{c_0}=\beta_{\rm v}M_0\frac{R}{R_c}\exp\!\left(-\frac{R^{2}}{2R_c^{2}}\right),
\quad
\frac{T}{T_0}=1-\tfrac{1}{2}(\gamma-1)M_0^{2}\beta_{\rm v}^{2}\exp\!\left(-\frac{R^{2}}{R_c^{2}}\right),
\label{eq:evc-ic}
\end{equation}
where $R$ is the distance from the vortex center, $u_\theta$ the tangential velocity, $T_0$ the background temperature, and $\beta_{\rm v}$ the vortex strength, unrelated to the $\beta_k$ of Eq.~\eqref{eq:weight}.
Homentropy gives $\rho/\rho_0=(T/T_0)^{1/(\gamma-1)}$ and the equation of state $\gamma p=(\rho/\rho_0)(T/T_0)$.
The peak density deficit is $\delta\rho/\rho_0\simeq M_0^{2}\beta_{\rm v}^{2}/2=5.0\times10^{-3}$ at the $\beta_{\rm v}=2$ used here (Figure~\ref{fig:cases}(b)), and the peak vortex Mach number is about 0.06, so the field remains smooth.
The grids are $n=64,128,256$, i.e.\ 3.2, 6.4, and 12.8 points per $R_c$.

\begin{figure}[t]
\centering
\includegraphics[width=\columnwidth]{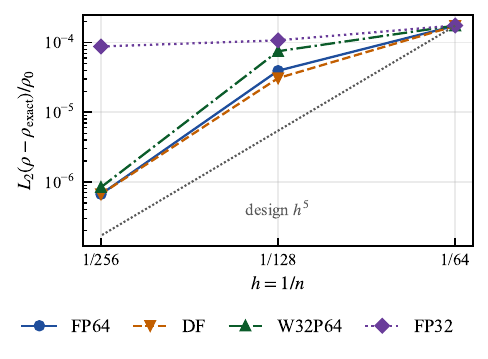}
\caption{Grid convergence on Euler vortex advection at WENO5, $\beta_{\rm v}=2$: the normalized density error $L_2(\rho-\rho_{\rm exact})/\rho_0$ after one period against $h=1/n$.
The dotted line is the slope of the design order $h^5$.}
\label{fig:evc-conv}
\end{figure}

Figure~\ref{fig:evc-conv} shows the results.
At $n=64$ the four placements agree within 1.3\% at $1.730$--$1.752\times10^{-4}$.
The local order, the slope of $\log L_2$ against $\log h$ between successive grids, is 5.88 for FP64 and 5.52 for DF emulation from $n=128$ to 256, near the design order 5; phase error remaining at $n=128$ and the time step limit this estimate to about one digit.
At this time step the W32P64 errors are 1.91 and 1.26 times those of FP64 at $n=128$ and 256.
We quote no local order for W32P64, because its $n=256$ error depends on the time step.
The all-FP32 reconstruction stops converging at a local order of 0.29; its $n=256$ error of $8.8\times10^{-5}$ is 1.8\% of the peak density deficit, 131 times that of FP64 and 104 times that of W32P64.
At WENO5 and this time step, then, moving only the weights to FP32 costs a factor of 1.26 in the $n=256$ error, whereas moving the whole reconstruction to FP32 costs a factor of 131.
This is consistent with the observation of Field et al.~\cite{field2023gpu} that the weights need less precision than the polynomial values.

\subsection{The solver time step on the roofline}
\label{sec:roof}

\begin{figure*}[t]
\centering
\includegraphics[width=\textwidth]{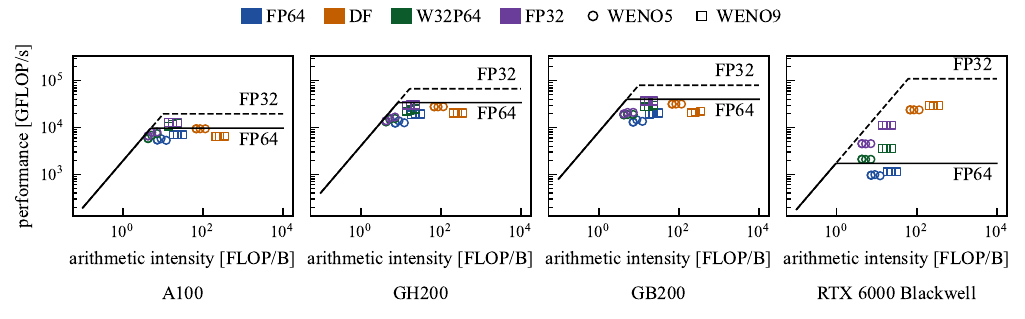}
\caption{The four stages of one classical Runge--Kutta step of the one-dimensional solver at WENO5 (circles) and WENO9 (squares), $n=2^{24}$, on the roofline of each GPU; colors distinguish the placements.
The intensity is FP64 plus FP32 FLOP (an FMA as two) per DRAM byte, and the ceilings are published peaks (solid: DRAM, FP64; dashed: FP32), the FP64 one of RTX 6000 Blackwell being FP32 over 64.}
\label{fig:roof}
\end{figure*}

Figure~\ref{fig:roof} places the four stages of one Runge--Kutta step of the one-dimensional solver on the roofline of each GPU.
Each stage moves at least four times the largest L2 cache, and the measured DRAM traffic is within 0.4\% of the 408 bytes per point and step that the stages must move.
On A100, GH200, and GB200, W32P64 achieves a speedup of 1.61--2.36$\times$ over FP64 at WENO5 and WENO9, whereas DF emulation makes it 4.0--12.3$\times$ slower.
On RTX 6000 Blackwell, DF emulation achieves a speedup of 2.27--2.63$\times$ and W32P64 a speedup of 3.71--4.18$\times$, against 4.47--5.92$\times$ for the W32P64 kernel alone (Table~\ref{tab:overview}).

The WENO step is not necessarily memory-bound: whether it is depends on the scheme order and the placement.
We compare the time of each step with the longest of its FP64 FLOP at the FP64 peak, its FP32 FLOP at the FP32 peak, and its DRAM bytes at the peak bandwidth.
Among FP64, W32P64, and DF emulation, only W32P64 at WENO5 on A100 and GH200 runs within twice its DRAM time (1.43--1.52 times).
FP64 runs at 1.42--1.87 times its FP64-pipe time on A100 and RTX 6000 Blackwell, and DF emulation at 2.10--4.67 times its FP32-pipe time on every GPU.
Register count, and hence occupancy, differs by placement and was not equalized: on RTX 6000 Blackwell, FP64 uses 56--76 registers at 50--66\% occupancy and W32P64 37--58 at 66--83\%.
 \section{Conclusion}
\label{sec:conclusion}

We investigated whether idle computing units can be utilized by concurrently executing instructions of different precisions on the same SM and, if so, whether this technique can be applied to a practical CFD kernel.

A100 shows both an inter-kernel throughput gain and intra-kernel overlap; GH200 and GB200 show neither.
On RTX 6000 Blackwell, the gain appears between kernels against FFMA and, against DF emulation, once the issue density of the DF kernel is reduced, but no overlap appears within a single kernel with FP32 weights and FP64 polynomials (W32P64).
On the GPUs evaluated here, intra-kernel overlap appeared only with an inter-kernel throughput gain, but the gain does not imply it.
On RTX 6000 Blackwell, even with free issue bandwidth, the added FP32
instructions raise register usage and lower occupancy, which may leave too few eligible warps to sustain the overlap.
Moreover, the issue-slot counters are consistent with the lane counts on A100, GH200, and GB200 but not on RTX 6000 Blackwell, so whether free slots can actually be used must be measured on each GPU.

The main findings are as follows.

\begin{enumerate}
\item In the microbenchmark, A100 sustains about 90\% of the standalone throughput on both the FP32 and FP64 sides with pure kernels, reducing wall time by 44.0\%, and gains $3.8$\% over native FP64 alone 
when running concurrently with a DF emulation kernel. 
RTX 6000 Blackwell gains $6.7$\% over DF emulation alone 
when the issue density of the DF kernel is reduced,
whereas on GH200
the combined throughput falls at least 40.2\% below native FP64 alone in every configuration, and on GB200 at least 47.6\%.
\item Inside the WENO-Z kernel, A100 gives an overlap indicator of $E=0.89$ and $0.84$ for WENO5 and WENO9, respectively, against $E\le 0.12$ on GH200 and GB200. 
  A forced-serialization control barely changes the run time, which is consistent with TLP but does not rule out intra-thread ILP.
\item On RTX 6000 Blackwell, the native FP64 pipe is the bottleneck, and applying DF emulation to the whole reconstruction reaches a 2.51--3.94$\times$ speedup over native FP64. 
Converting only the weight computation to DF emulation achieves only 2.39--3.05$\times$ speedup; 
the remaining gain comes from also moving the FP64 instructions left in the polynomial computation onto the FP32 pipe.
\item The fastest placement does not always preserve the design order of accuracy. For the one-dimensional entropy wave, 
W32P64 retains 4.91 for WENO5 but drops to 1.78 for WENO7 and $-2.94$ for WENO9; for WENO9 with $n = 64$, its error is seven orders of magnitude larger than that of FP64.
DF emulation stays close to the design order in all three cases on the grids tested. For the two-dimensional Euler vortex with WENO5, at the time step used, the W32P64 error is 1.26$\times$ that of FP64 at $n=256$, while the
  all-FP32 placement stops converging (local order 0.29, with an error 104$\times$ that of W32P64).
\end{enumerate}

To realize the performance benefits of mixed-precision computation and DF emulation in CFD solvers, effective redistribution of computational resources is essential. Although NVIDIA GPUs generally offer limited FP64 throughput, increasing the instruction issue bandwidth could enable the concurrent execution of instructions with different precisions, thereby utilizing otherwise idle execution units and improving overall performance.

Future work will investigate the impact of W32P64 on high-order turbulence statistics and shock-containing flows, as well as extend the proposed resource redistribution strategy to FP16 and integer execution units.

Ultimately, our goal is to develop novel CFD algorithms that fully exploit the available computational and memory resources of each GPU architecture.

\begin{acks}
This research was supported by the Collaborative Research Project on Computer Science with High-Performance Computing at Nagoya University and the Core Facility of Okayama University (CFPOU Engineering 001), which provided computational resources. In addition, this work received financial support from JST SPRING, Japan (Grant Number JPMJSP2126), and the Public Interest Incorporated Foundation Ohmoto Ikueikai. During this work, the authors used Claude Code, Claude, Codex, and ChatGPT for code development and manuscript writing.
\end{acks}
 
\bibliographystyle{ACM-Reference-Format}

\begin{thebibliography}{14}

\ifx \showCODEN    \undefined \def \showCODEN     #1{\unskip}     \fi
\ifx \showISBNx    \undefined \def \showISBNx     #1{\unskip}     \fi
\ifx \showISBNxiii \undefined \def \showISBNxiii  #1{\unskip}     \fi
\ifx \showISSN     \undefined \def \showISSN      #1{\unskip}     \fi
\ifx \showLCCN     \undefined \def \showLCCN      #1{\unskip}     \fi
\ifx \shownote     \undefined \def \shownote      #1{#1}          \fi
\ifx \showarticletitle \undefined \def \showarticletitle #1{#1}   \fi
\ifx \showURL      \undefined \def \showURL       {\relax}        \fi
\providecommand\bibfield[2]{#2}
\providecommand\bibinfo[2]{#2}
\providecommand\natexlab[1]{#1}
\providecommand\showeprint[2][]{arXiv:#2}
\makeatletter
\@ifundefined{NAT@parse@date}{}{\let\NAT@parse@date@orig\NAT@parse@date}
\@ifundefined{NAT@parse@date}{}{\def\NAT@parse@date#1#2#3#4#5#6@@{\NAT@parse@date@orig#1#2#3#4#5#6@@\def\NAT@tempyear{0000}\def\NAT@tempexlab{{?}}\ifx\NAT@year\NAT@tempyear\ifx\NAT@exlab\NAT@tempexlab\def\NAT@date{[n.\,d.]}\else\edef\NAT@date{[n.\,d.]\NAT@exlab}\fi\fi}}
\makeatother

\bibitem[Borges et~al\mbox{.}(2008)]{borges2008improved}
\bibfield{author}{\bibinfo{person}{Rafael Borges}, \bibinfo{person}{Monique
  Carmona}, \bibinfo{person}{Bruno Costa}, {and} \bibinfo{person}{Wai~Sun
  Don}.} \bibinfo{year}{2008}\natexlab{}.
\newblock \showarticletitle{An improved weighted essentially non-oscillatory
  scheme for hyperbolic conservation laws}.
\newblock \bibinfo{journal}{\emph{J. Comput. Phys.}} \bibinfo{volume}{227},
  \bibinfo{number}{6} (\bibinfo{year}{2008}), \bibinfo{pages}{3191--3211}.
\newblock

\bibitem[Castro et~al\mbox{.}(2011)]{castro2011high}
\bibfield{author}{\bibinfo{person}{Marcos Castro}, \bibinfo{person}{Bruno
  Costa}, {and} \bibinfo{person}{Wai~Sun Don}.}
  \bibinfo{year}{2011}\natexlab{}.
\newblock \showarticletitle{High order weighted essentially non-oscillatory
  WENO-Z schemes for hyperbolic conservation laws}.
\newblock \bibinfo{journal}{\emph{J. Comput. Phys.}} \bibinfo{volume}{230},
  \bibinfo{number}{5} (\bibinfo{year}{2011}), \bibinfo{pages}{1766--1792}.
\newblock

\bibitem[Chen et~al\mbox{.}(2026)]{chen2026enabling}
\bibfield{author}{\bibinfo{person}{Yanxiang Chen}, \bibinfo{person}{Pablo de
  Oliveira~Castro}, \bibinfo{person}{Paolo Bientinesi}, \bibinfo{person}{Niclas
  Jansson}, {and} \bibinfo{person}{Roman Iakymchuk}.}
  \bibinfo{year}{2026}\natexlab{}.
\newblock \showarticletitle{Enabling mixed-precision in spectral element
  codes}.
\newblock \bibinfo{journal}{\emph{Future Generation Computer Systems}}
  \bibinfo{volume}{174} (\bibinfo{year}{2026}), \bibinfo{pages}{107990}.
\newblock

\bibitem[Dekker(1971)]{dekker1971floating}
\bibfield{author}{\bibinfo{person}{Theodorus~Jozef Dekker}.}
  \bibinfo{year}{1971}\natexlab{}.
\newblock \showarticletitle{A floating-point technique for extending the
  available precision}.
\newblock \bibinfo{journal}{\emph{Numer. Math.}} \bibinfo{volume}{18},
  \bibinfo{number}{3} (\bibinfo{year}{1971}), \bibinfo{pages}{224--242}.
\newblock

\bibitem[Field et~al\mbox{.}(2023)]{field2023gpu}
\bibfield{author}{\bibinfo{person}{Scott~E Field}, \bibinfo{person}{Sigal
  Gottlieb}, \bibinfo{person}{Zachary~J Grant}, \bibinfo{person}{Leah~F
  Isherwood}, {and} \bibinfo{person}{Gaurav Khanna}.}
  \bibinfo{year}{2023}\natexlab{}.
\newblock \showarticletitle{A GPU-accelerated mixed-precision WENO method for
  extremal black hole and gravitational wave physics computations}.
\newblock \bibinfo{journal}{\emph{Communications on Applied Mathematics and
  Computation}} \bibinfo{volume}{5}, \bibinfo{number}{1}
  (\bibinfo{year}{2023}), \bibinfo{pages}{97--115}.
\newblock

\bibitem[Hida et~al\mbox{.}(2001)]{hida2001algorithms}
\bibfield{author}{\bibinfo{person}{Yozo Hida}, \bibinfo{person}{Xiaoye~S Li},
  {and} \bibinfo{person}{David~H Bailey}.} \bibinfo{year}{2001}\natexlab{}.
\newblock \showarticletitle{Algorithms for quad-double precision floating point
  arithmetic}. In \bibinfo{booktitle}{\emph{Proceedings 15th IEEE Symposium on
  Computer Arithmetic. ARITH-15 2001}}. IEEE, \bibinfo{pages}{155--162}.
\newblock

\bibitem[Jiang and Shu(1996)]{jiang1996efficient}
\bibfield{author}{\bibinfo{person}{Guang-Shan Jiang} {and}
  \bibinfo{person}{Chi-Wang Shu}.} \bibinfo{year}{1996}\natexlab{}.
\newblock \showarticletitle{Efficient implementation of weighted ENO schemes}.
\newblock \bibinfo{journal}{\emph{J. Comput. Phys.}} \bibinfo{volume}{126},
  \bibinfo{number}{1} (\bibinfo{year}{1996}), \bibinfo{pages}{202--228}.
\newblock

\bibitem[Karp et~al\mbox{.}(2025)]{karp2025effects}
\bibfield{author}{\bibinfo{person}{Martin Karp}, \bibinfo{person}{Ronith
  Stanly}, \bibinfo{person}{Timofey Mukha}, \bibinfo{person}{Luca Galimberti},
  \bibinfo{person}{Siavash Toosi}, \bibinfo{person}{Hang Song},
  \bibinfo{person}{Lisandro Dalcin}, \bibinfo{person}{Saleh Rezaeiravesh},
  \bibinfo{person}{Niclas Jansson}, \bibinfo{person}{Stefano Markidis},
  {et~al\mbox{.}}} \bibinfo{year}{2025}\natexlab{}.
\newblock \showarticletitle{Effects of lower floating-point precision on
  scale-resolving numerical simulations of turbulence}.
\newblock \bibinfo{journal}{\emph{J. Comput. Phys.}} (\bibinfo{year}{2025}),
  \bibinfo{pages}{114600}.
\newblock

\bibitem[Kashi et~al\mbox{.}(2026)]{kashi2026mixed}
\bibfield{author}{\bibinfo{person}{Aditya Kashi}, \bibinfo{person}{Hao Lu},
  \bibinfo{person}{Wesley Brewer}, \bibinfo{person}{David Rogers},
  \bibinfo{person}{Michael Matheson}, \bibinfo{person}{Mallikarjun Shankar},
  {and} \bibinfo{person}{Feiyi Wang}.} \bibinfo{year}{2026}\natexlab{}.
\newblock \showarticletitle{Mixed-precision numerics in scientific
  applications: survey and perspectives}.
\newblock \bibinfo{journal}{\emph{The Journal of Supercomputing}}
  \bibinfo{volume}{82}, \bibinfo{number}{5} (\bibinfo{year}{2026}),
  \bibinfo{pages}{287}.
\newblock

\bibitem[Liu et~al\mbox{.}(1994)]{liu1994weighted}
\bibfield{author}{\bibinfo{person}{Xu-Dong Liu}, \bibinfo{person}{Stanley
  Osher}, {and} \bibinfo{person}{Tony Chan}.} \bibinfo{year}{1994}\natexlab{}.
\newblock \showarticletitle{Weighted essentially non-oscillatory schemes}.
\newblock \bibinfo{journal}{\emph{J. Comput. Phys.}} \bibinfo{volume}{115},
  \bibinfo{number}{1} (\bibinfo{year}{1994}), \bibinfo{pages}{200--212}.
\newblock

\bibitem[{NVIDIA}(2020)]{nvidia2020ampere}
\bibfield{author}{\bibinfo{person}{{NVIDIA}}.} \bibinfo{year}{2020}\natexlab{}.
\newblock \bibinfo{booktitle}{\emph{{NVIDIA A100 Tensor Core GPU}
  Architecture}}.
\newblock \bibinfo{type}{White Paper}. \bibinfo{institution}{NVIDIA
  Corporation}.
\newblock
\urldef\tempurl \url{https://images.nvidia.com/aem-dam/en-zz/Solutions/data-center/nvidia-ampere-architecture-whitepaper.pdf}
\showURL{\tempurl}

\bibitem[{NVIDIA}(2023)]{nvidia2022hopper}
\bibfield{author}{\bibinfo{person}{{NVIDIA}}.} \bibinfo{year}{2023}\natexlab{}.
\newblock \bibinfo{booktitle}{\emph{{NVIDIA H100 Tensor Core GPU}
  Architecture}}.
\newblock \bibinfo{type}{White Paper}. \bibinfo{institution}{NVIDIA
  Corporation}.
\newblock
\urldef\tempurl \url{https://resources.nvidia.com/en-us-hopper-architecture/nvidia-h100-tensor-c}
\showURL{\tempurl}

\bibitem[{NVIDIA Corporation}(2025)]{nvidia2025cudapg}
\bibfield{author}{\bibinfo{person}{{NVIDIA Corporation}}.}
  \bibinfo{year}{2025}\natexlab{}.
\newblock \bibinfo{booktitle}{\emph{{CUDA C++} Programming Guide}}.
\newblock NVIDIA Corporation.
\newblock
\urldef\tempurl \url{https://docs.nvidia.com/cuda/archive/13.0.0/cuda-c-programming-guide/index.html}
\showURL{\tempurl}
\newblock
\shownote{Version 13.0}.

\bibitem[Sikl{\'o}si et~al\mbox{.}(2025)]{siklosi2025reduced}
\bibfield{author}{\bibinfo{person}{B{\'a}lint Sikl{\'o}si},
  \bibinfo{person}{Pushpender~K Sharma}, \bibinfo{person}{David~J Lusher},
  \bibinfo{person}{Istv{\'a}n~Z Reguly}, {and} \bibinfo{person}{Neil~D
  Sandham}.} \bibinfo{year}{2025}\natexlab{}.
\newblock \showarticletitle{Reduced and mixed precision turbulent flow
  simulations using explicit finite difference schemes}.
\newblock \bibinfo{journal}{\emph{Future Generation Computer Systems}}
  (\bibinfo{year}{2025}), \bibinfo{pages}{108111}.
\newblock

\end{thebibliography}

 \end{document}